# Apparatus for Seebeck coefficient measurement of wire, thin film and bulk materials in the wide temperature range (80 – 650 K)

Ashish Kumar[1,*], Ashutosh Patel[2], Saurabh Singh[3], K. Asokan[1], D. Kanjilal[1]

[1]*Inter-University Accelerator Centre, New Delhi, 110067, India*

[2]*Department of Mechanical Engineering, IISc Bangalore, 560012, India*

[3]*Toyota Technological Institute, Hisakata 2-12-1, Tempaku, Nagoya 468-8511, Japan.*

*email address: ashish@iuac.res.in, dr.akmr@gmail.com

A Seebeck coefficient measurement apparatus has been designed and developed, which is very effective for accurate characterization of different type of samples in a wide temperature range (80 – 650K) simultaneously covering low as well as high temperature regime. Reducing complexity of technical design of sample holder and data collections has always been challenging to implement in a single instrument when samples are in different geometrical shape and electronic structure. Our unique design of sample holder with pressure probes covers measurements of different samples shapes (wires, thin films and pellets) as well as different resistivity ranges (metals, semiconductors and insulators). It is suitable for characterization of different samples sizes (3–12 mm). A double heater configuration powered by a dual channel source meter is employed for maintaining a desired constant temperature difference across the sample for the whole temperature range. Two *K*-type thermocouples are used for simultaneously reading of temperatures and Seebeck voltages by utilizing different channels of a multichannel digital multimeter. Calibration of the system has been carried out using constantan, chromel and alumel materials and recorded data is found to be very accurate and consistent with earlier reports. The Seebeck coefficients of standard samples of constantan (wire) and GaN (thin film) have been reported, which shows the measurement capability of designed setup with versatile samples.

## I. INTRODUCTION

Thermoelectric materials are known to be as electronic materials capable of converting thermal gradients into electrical voltage and vice-versa [1]. Former property has resulted into applications of power generation in deep space exploration modules [2], waste heat recovery in vehicles [3], wireless remote sensing [4,5] while the latter found uses in cooling delicate electronics, small scale refrigeration [6]. One inherent advantage of thermoelectric devices is that being a solid state device, they are easily scalable and need low maintenance over time. This has led to a growing interest in the development of better thermoelectric materials where a relatively small increment of 20% over state-of-art may lead to significant commercial success. Accurate measurement of parameters has critical importance as in various reports of inaccurate measurement often leads to misinterpretation, boosted values of Seebeck coefficients [7-9] and thermal conductivity [10,11]. Various excellent review articles have been reported on thermoelectric instrumentation and measurement, new modes and data treatment [12-15]. Seebeck coefficient is essentially a relative measurement between two materials and hence particularly difficult to measure. For low temperatures (10–390 K), a standard reference material (NIST SRM 3451) is used but no such suitable standard reference is available for the high-temperature range [16]. Constantan, chromel and thermocouple alloys have well-reported data for low Seebeck coefficient range. Though it depends on the technique and design of instruments used but a typical variation of 5–8 % in measured values of Seebeck coefficient at room temperature is observed by different groups[16-20]. Conceptually, the Seebeck coefficient is defined as the ratio of emf developed to the applied temperature difference across the sample. There are different methods of measurements for two parameters (thermo-emf and temperature) have been used. For example, in the slope method, the temperature is varied about a fixed average temperature and the slope of voltage *vs*. temperature gradient curve is used for calculating the Seebeck coefficient. In the single point measurement method, the ratio of voltage corresponding to a single temperature difference is used. If the desired temperature gradient is stabilized steadily before each reading [12,21], the method is called steady state measurement while for quasi-steady state method, voltage is continuously recorded as the temperature difference is changed slowly [21-23]. Few important tips can be emphasized here, *e.g*. a good thermal and electrical (ohmic) contact should be ensured between sample and probes for accurate measurement. The temperature gradient should also be kept optimum (4–20 K) for complete measurement range. The very small temperature gradient may result in inaccurate

Seebeck coefficient. The large temperature gradient is very difficult to maintain for all kind of samples. Various challenges of instrument design [24] and data analysis have been reviewed recently [25]. Characterization of Seebeck coefficient in wide temperature range including low and high temperature range in single measurement system is still a great challenge. An accurate measurement with low cost system is highly desirable to explore the thermoelectric materials in wide temperature range. Thus, we have designed the sample holder which is very suitable and easy to fit in the cryostat, and dedicated automation of setup using Labview program is done for controlling and colleting the accurate data with desired sample gradient and sample temperature.

In the present work, we have developed a steady state Seebeck coefficient measurement setup for wide temperature range (80–650 K). A low-temperature vacuum cryostat with very simple dipstick geometry has been designed for keeping sample holder assembly. The sample holder can be used for different sample shapes (wire, thin film, bulk, irregular and symmetric) and sizes (3–12 mm). Two heaters with live control of the hot and cold side temperature helps to maintain the temperature difference within permissible limit throughout the measurement. The compact design, controlled temperature difference, and capability to measure from 80 K to 650 K in a single run are the main silent features of this setup Detailed design and novelty aspects have been discussed further in the following sections.

## II. THEORY AND METHODOLOGY

The Seebeck coefficient measurement is basically carried out by using two different methods, classified as integral and differential modes. In the former method, cold junction temperature ($T_C$) is kept constant and the temperature of hot junction ($T_H$) is varied. The Seebeck voltage ($\Delta V$) measured thus can be plotted as a function of hot junction temperature ($T_H$). Advantage of this method is that a large Seebeck voltage is generated due to the large temperature gradient across the sample this minimizing errors due to any spurious fluctuations in the circuit [26]. Limitations of this technique lie in the complexity of design in keeping cold junction temperature fixed, and for nondegenerate semiconductors and insulators [27,28].

The second method i.e. differential method is widely used and is given by

$$S_S = -\frac{\Delta V}{\Delta T} + S_W \quad (1)$$

Here $S_S$, $S_W$ and $\Delta V$ are absolute Seebeck coefficients of sample, wire and measured voltage, respectively. This method doesn't require specific cooling arrangements and can be used for all materials [29-33].

Generally, low Seebeck coefficient materials such as copper, niobium, and platinum are used as to measure the thermoelectric voltage across the sample. Ideally, the temperatures and voltages should be read from exactly same points which requires the thermocouples and copper wire (for voltage) should be fixed at same points of sample requiring special design considerations [34]. Boor *et. al.* [35,36] suggested a solution for the above problem by modifying the conventional equation in a new form:

$$S_S = -\frac{V_{neg}}{V_{pos} - V_{neg}} S_{TC} + S_{neg} \quad (2a)$$

Here $V_{pos}$ and $V_{neg}$ are Seebeck voltages measured by positive (chromel) legs and negative (alumel) legs of thermocouples wires, respectively. $S_{TC}$ and $S_{neg}$ are Seebeck coefficients of thermocouple and its negative leg, respectively. Similar equation using the positive legs of thermocouples can also be written as where $S_{neg}$ is replaced by $S_{pos}$.

$$S_S^{"} = -\frac{V_{pos}}{V_{neg} - V_{pos}} S_{TC} + S_{pos} \quad (2b)$$

Where $S_{pos}$ is the Seebeck coefficient of positive leg of the thermocouple. This approach eliminates the requirement of additional connecting wires to measure the thermoelectric voltage across the sample. Singh *et al.* have implemented this strategy using single heater setup, and two thermocouples to measure thermo-emf voltages [37]. This removes any errors which may generate when Seebeck voltage and temperature are not measured exactly at the same point on the sample apart from the advantage of simplified design. Additionally, the spurious thermal offset voltages from the system are cancelled out [35,36]. As Seebeck coefficient varies with temperature, finding $S_{pos}$, $S_{neg}$ and $S_{TC}$ for a temperature range from $T_C$ to $T_H$ at every data point makes the measurement procedure complex [40]. For small $\Delta T$, Kolbe et. al. [38] and Boor et. al. [35,36] approximated values of $S_{TC}$ and $S_{neg}$ as

$$S_{TC}(T_C, T_H) \approx S_{TC}(\bar{T}) \qquad (3)$$

$$S_{neg}(T_C, T_H) \approx S_{neg}(\bar{T}) \qquad (4)$$

For linear varying Seebeck coefficients of cromel, alumel and $K$-type thermocouple in the temperature range of $T_C$ to $T_H$, the above approximation will work perfectly, which is difficult to meet. The deviation from actual will increase as the temperature difference and non-linearity increases. The chance of inaccuracy in the finding of $S_{pos}$, $S_{neg}$ and $S_{TC}$ can be eliminated by implementing the ethodology discussed by Patel et. al. [40]. Further, to calculate the Seebeck coefficients of thermocouple and the wire (negative leg) we used a modified Boor's method as discussed below [39,40]. The general expression for Seebeck voltage for a sample can be written as

$$V_S(T_C, T_H) = -\int_{T_C}^{T_H} S_S(T) \, dT \qquad (5)$$

where $T_H$, $T_C$ and $S_S$ are the temperatures of the hot junction, cold junction and Seebeck coefficient as a function of temperature. If connecting wires are used for recording the Seebeck voltage of sample, two ends of connecting wires are at different temperature i.e. free ends are at temperature $T_{cj}$ and other ends are $T_H$, (or $T_C$). So it also adds its own Seebeck voltage in measured values. Thus the measured voltage is the sum of connecting wires voltages at both ends of the sample and Seebeck voltage of sample.

$$V_m(T_C, T_H) = V_{wc} + V_S + V_{wh} \qquad (6)$$

Using equation (5) for writing the expression of $V_{wc}$ and $V_{wh}$ in equation (6) provides

$$V_m(T_C, T_H) = V_S - \int_{T_R}^{T_C} S_W(T)\,dT - \int_{T_H}^{T_R} S_W(T)\,dT$$

$$= V_S + \int_{T_C}^{T_R} S_W(T)\,dT + \int_{T_R}^{T_H} S_W(T)\,dT$$

$$= V_S + \int_{T_C}^{T_H} S_W(T)\,dT \qquad (7)$$

The second term in equation (7) represents net Seebeck voltage contribution from the wires (say $V_w$),

$$V_w(T_C, T_H) = -\int_{T_C}^{T_H} S_w(T)\,dT \qquad (8)$$

The effective Seebeck coefficient for wires operating between temperatures $T_H$ and $T_C$ can be written as

$$S_w(T_C, T_H) = -\frac{1}{T_H - T_C}\int_{T_C}^{T_H} S_w(T)\,dT \qquad (9)$$

If the negative legs of two thermocouples are used as connecting wires for reading Seebeck voltages, then $S_W$ can be written as $S_{neg}$. Similarly, the Seebeck coefficient for the thermocouple can also be written as

$$S_{TC}(T_C, T_H) = -\frac{1}{T_H - T_C}\int_{T_C}^{T_H} S_{TC}(T)\,dT \qquad (10)$$

Equations (9) and (10) have been used for calculation of Seebeck coefficients samples[39]. Advantage of the above equation is that they can be used for large temperature gradients ($\Delta T$) also.

### III. MEASUREMENT SETUP

### A. System Description

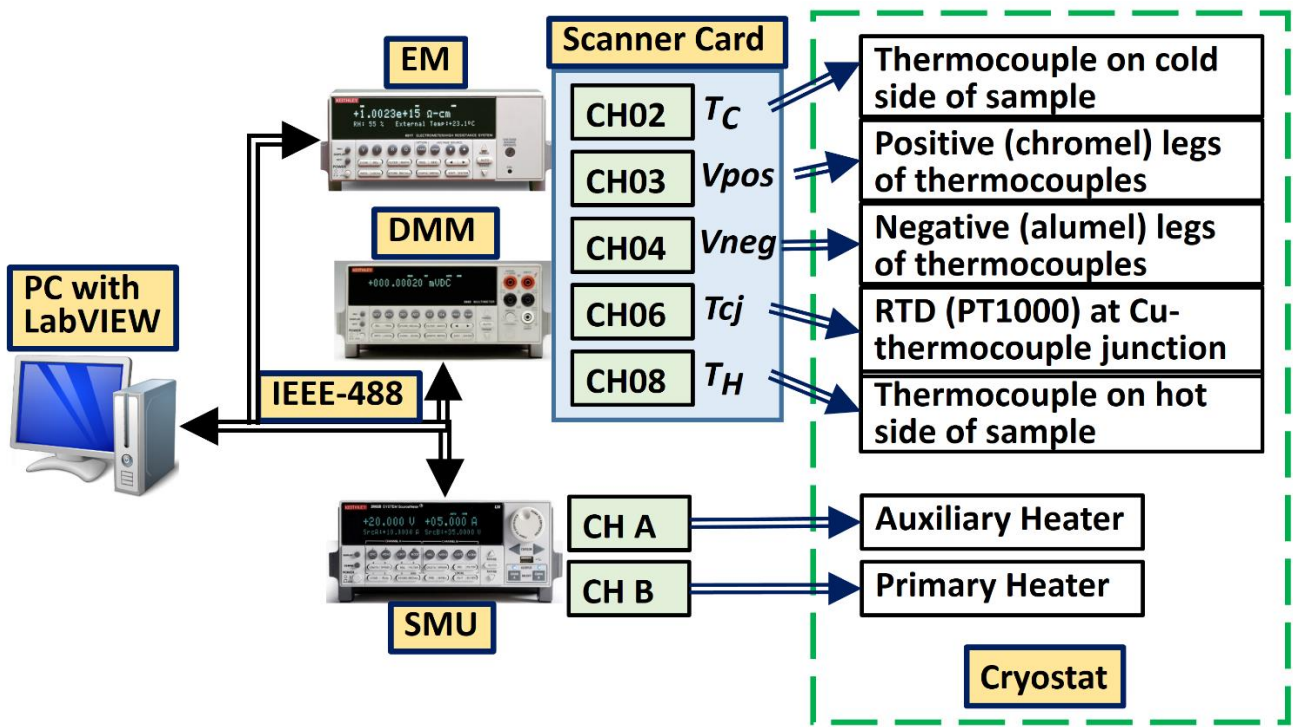

**Figure 1** Instrumentation schematic consist of three main components; a desktop, a DMM, an SMU and an optional electrometer (for high resistive sample) interfaced with LabVIEW.

A schematic diagram of the instrumentation is shown in figure 1. It consists of a digital multimeter (DMM) with 10 channel scanner card for simultaneous measurement of various signals. A dual channel source measure unit (SMU) is used to power the main and auxiliary heaters. Both the instruments are GPIB interfaced with a computer using LabVIEW software. Different channels of DMM have been connected to thermocouples and PT-1000 RTD. An electrometer (EM) is also connected as and when a high resistivity sample is needed to characterize. The high input impedance (> 200 T$\Omega$) of electrometer prevents loading and accurate measurement of thermo-emf is performed. $T_H$ and $T_C$ represent the temperature of the hot and cold junction of the sample measured by $K$-type (chromel - alumel) thermocouple, respectively. Voltages $V_{pos}$ and $V_{neg}$ are the Seebeck voltages read by chromel and alumel legs in two thermocouples. A temperature sensor (PT-1000 RTD) is mounted close to free end (copper thermocouple) junction for cold junction compensation ($Tcj$). All wires are connected to the cryostat via feed thru adapter mounted at the top. The sample holder design is shown in figure 2 where different components are labelled by numbers. The dipstick cryostat is made of hollow steel pipe having 20 mm diameter and one-meter length (1). Bottom of it has the sample holder (35 mm dia.) (2) having threads for a cup cover. This portion is submerged in liquid nitrogen dewar for low-temperature measurements. Two specially designed copper blocks (5) are fixed in holder with help of copper studs (3) and held close by a circular gypsum block (6) at the bottom. Gypsum has very low thermal conductivity. So it can maintain the temperature gradient between the copper blocks. Both the semi-circular copper blocks are isolated electrically from each other (~2 mm gap) and the main body by high-temperature ceramic washers (4). These copper blocks have been fitted with two identical commercial pencil heaters (11) (Lakeshore, 25 ohm). The sample is kept over these blocks. Two copper pressure probes (9) are mounted with rotation freedom (x-y plane) for adjusting irregular shape and size samples. Two polytetrafluoroethylenes coated *K*-type thermocouples (8) of 36 swg are mounted in these blocks through small drills. GaSn is used for making good thermal and electrical contact between thermocouple and copper probes as and when required for samples (pellets). A rotary pump is used for maintaining vacuum ($< 10^{-3}$ torr) in the cryostat.

The interfacing and data acquisition program has been written in LabVIEW code. The power supply (using SMU) of the main heater is controlled and increased systematically to achieve a certain temperature ($T_H$). The second channel of the SMU is for supplying power to the auxiliary heater to maintain temperature difference ($\Delta T$) within the permissible limit across the sample. The DMM starts acquiring data for different voltages only when a steady state of temperature is achieved. Different averaging schemes (multiple points or repetition) have been employed to minimize the error. The expression for $S_{TC}$ and $S_{neg}$ (equation (9) and (10)) has been implemented in LabVIEW program which requires polynomial coefficients of temperature dependent Seebeck coefficient function, $T_C$ and $T_H$. The polynomial coefficients have been taken from the literature [36,41] and fitted for 19-degree equation. The measured values of $V_{pos}$ and $V_{neg}$ along with the $S_{TC}$ and $S_{neg}$ are used in equation (2) to determine the sample Seebeck coefficient ($S_S$).

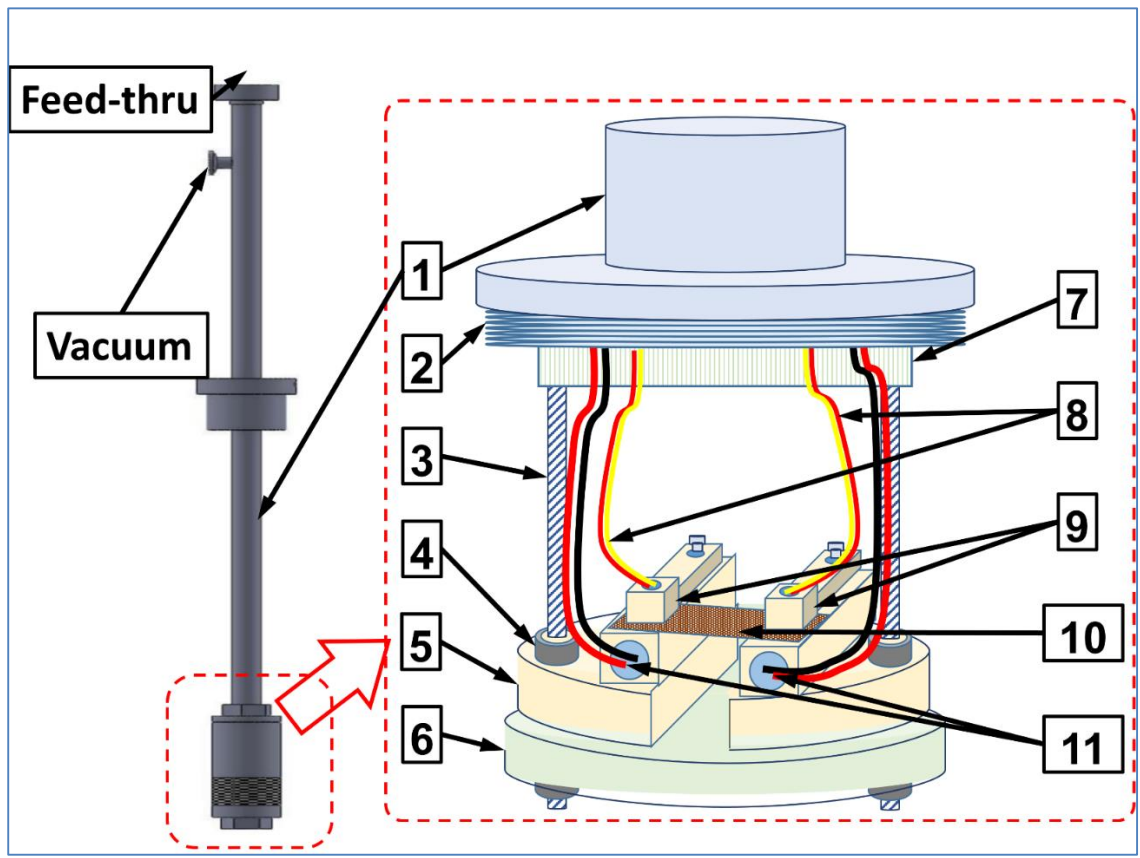

**Figure 2 Sample holder design**

## B. Procedure

The sample is mounted under the two copper pressure probes with or without the (In+Sb) eutectic layer for better thermal contact, depending upon sample requirement. After putting the SS cup holder, a low vacuum is created with a rotary pump and the dipstick cryostat is merged in liquid nitrogen dewar. All the input parameters and information ($\Delta T$, highest measurement temp., sample and file name, etc.,) is saved in the program before the run. Once the lowest steady state temperature is recorded (77 K) by both the thermocouples, the primary heater power is increased in small steps ensuring a temperature gradient ($\Delta T$) of the desired value (~5 K). Measurement methodology adopted ensures steady state equilibrium condition before each data recording. Various parameters ($T_H$, $T_C$, $V_{pos}$, $V_{neg}$, $R_{cj}$, data logging time, and heater powers) are recorded very fast (micro to mill seconds) using DMM scanner card switching option and resultant output quantities ($S_{TC}$, $S_{neg}$, $S_{pos}$) are calculated and saved in a .xml file in real time. After this, the program increases the power of the primary heater for next higher temperature measurement. Auxiliary heater power is automatically adjusted to keep $\Delta T$ constant. Once the highest temperature is reached by the sample, measurement is stopped or the cooling cycle starts depending upon the input conditions.

## IV. RESULTS AND DISCUSSION

Instrument calibration was performed by using constantan, chromel, alumel (wires) and GaN (thin film) samples for the whole temperature range. The repeatability and reproducibility of data were verified by remounting the same sample before each successive measurements. Figure 3 shows the temperature recorded by thermocouple (type K) temperatures at hot ($T_H$) and cold ($T_C$) side of sample with respect to the mean or average temperature (= ½ *($T_H$ +$T_C$)) for a fixed input value (5 K) of $\Delta T$. The measurements were performed in steady-state conditions where temperature equilibrium reached an average 50 minutes for successive data points. The temperature steps for each successive measurements are around 10 Kelvin. It can be pointed out that neither the equilibrium (steady state) time nor the temperature steps are predefined or user controlled. However, the temperature difference across the two ends of the sample ($\Delta T$) can be kept roughly fixed around the desired value (5 K in this graph). These parameters are dependent on steady state equilibrium conditions which itself depends on the thermal conductivity and dimensions of the given sample. The linearity of two curves over the whole temperature ranges show the stability of heater power control and measurements. The inset of figure 3 shows the variation of the temperature gradient generated ($\Delta T$) as a function of mean temperature (*T*). The temperature at the free end of thermocouples (at the feed-thru junction), *Tcj*, is also plotted. This temperature is an indicator of a stable temperature environment of the lab and is used for thermocouple cold junction compensation correction.

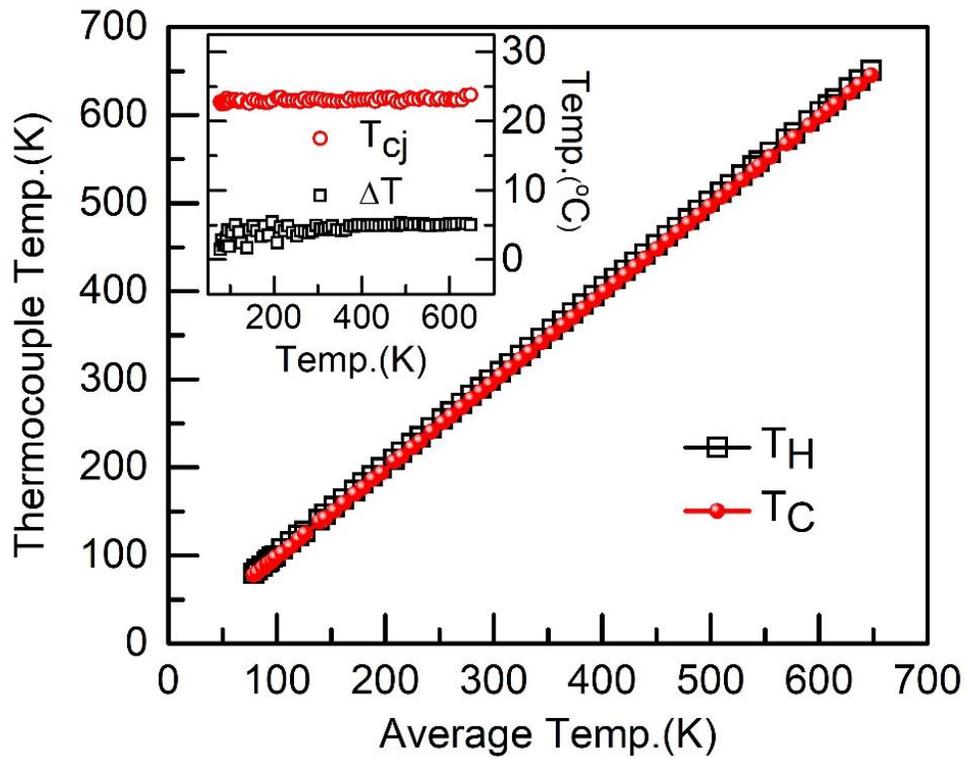

**Figure 3** Variation of hot and cold side temperatures (T$_H$ and T$_C$) with mean temperature with sample mounted condition. (Variation in temperature gradient (∆*T*) and free end junction temperature (*Tcj*) with Average Temp is shown in the inset)

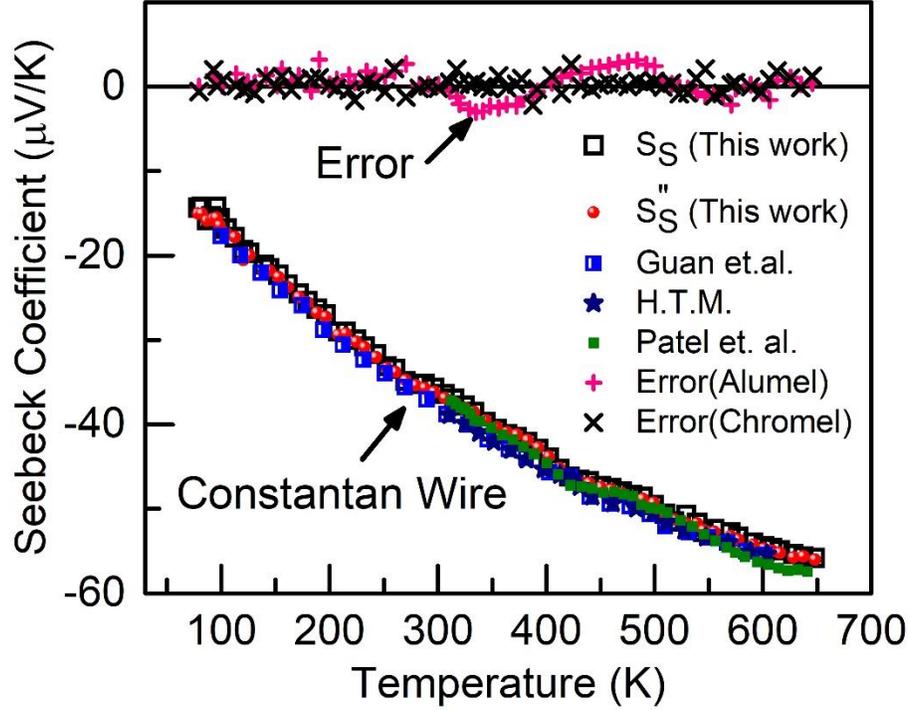

**Figure 4** Variation of Seebeck coefficient versus mean temperature of constantan sample, where the data corresponding to $S_S$ and $S_S"$ of present work, the data are taken from Handbook of temperature measurement [41], Patel *et. al* [40], and by Guan *et. al.* [23]. The curve named error are the noise voltages recorded by the alumel and chromel legs of thermocouples when alumel and chromel samples were mounted.

Figure 4 shows the Seebeck coefficients of constantan wire as a function of average temperature measured in this work and data available in the literature. $S_S$ is the Seebeck coefficient constantan wire recorded as a function of temperature by negative legs of thermocouples as given by equation 2(a). Similarly, $S_S"$ corresponds to Seebeck voltage measured by positive legs of thermocouples and is defined by equation 2(b). The maximum difference between former and latter over the whole temperature range is less than ±1 µV/K, which indicates a better thermal contact between thermocouple with probes and sample and the measurement is rather reliable. This data has been compared with the available literature. Guan *et. al.* [23] have reported the Seebeck coefficient values of constantan metal for the temperature range 100 – 600 K measured by two heater geometry setup and small $\Delta T$. The value of $S_S$ at room temperature (300 K) is 35.65 µV/K (and 36.1 µV/K for $S_S"$ in figure 4) which is in close approximation of reported standard value (35.0 µV/K) taken from Handbook of temperature measurement [41]. Patel *et. al.* [40] have also measured Seebeck coefficient of constantan from 300 K to 600 K on their system which was a single heater operated design with large $\Delta T$. Measurement of the absolute value of noise floor in any system is a trickier exercise. By mounting an alumel sample and recording its Seebeck voltage through negative legs of thermocouple (alumel wires) should ideally give zero voltages as the circuit is completed by the same material. Any nonzero voltages in this configuration give maximum instrumental noise for all temperatures. In our system, we measured it by recording Seebeck voltage of alumel and chromel wires by using negative and positive legs, respectively as shown by curves "Érrors" in figure 4). Maximum absolute values for both curves is the whole temperature range is less than ±2.2 µV/K.

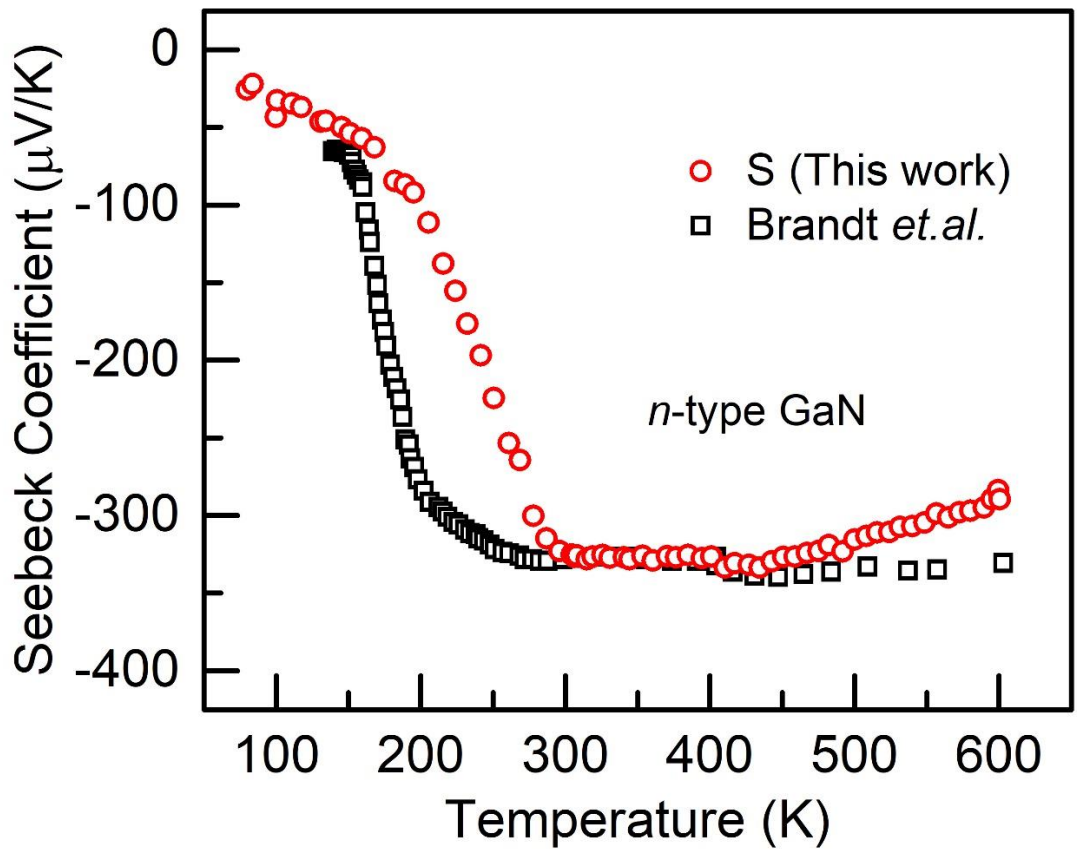

**Figure 5 Variation of Seebeck coefficient versus mean temperature of the *n*-type GaN sample measured in present work and literature.**

Measurement of the Seebeck coefficient of *n*-type GaN is carried out. GaN thin films (3.5 μm) deposited on sapphire substrate by metal organic chemical vapour deposition technique has been utilized for the present study. GaN is a wide band gap material (3.4 eV), is expected to have a high Seebeck coefficient. Also, it has a high breakdown voltage attributing its high operational temperature. Implementing on-chip thermoelectric devices made from GaN in the semiconductor industry has greater potential in Seebeck and Peltier effect based applications in small scale smart devices. Figure 5 shows the variation of the Seebeck coefficient of *n*-type GaN measured over 80 – 650 K temperature range. Comparing with data reported by Brandt *et. al.* [42], it can be observed that *S* increases in magnitude with temperature up to 300 K and saturates for higher temperatures. The Seebeck coefficients in both the data have similar trends but the maximum magnitude and saturation temperature are different. This may be due to the different growth methods of films adopted and different carrier concentration profiles. The films in the reported data [42] are grown by molecular beam epitaxy (MBE) and has higher carrier concentration ($2\times10^{18}$ /cm$^3$) compared to our sample (~$4\times10^{16}$ /cm$^3$)[43,44]. The detailed mechanism of the observed trend of Seebeck coefficient in GaN is affected by defects present and may also involve secondary transport mechanisms [42,45]. Detailed analysis can be part of a separate study.

**V. CONCLUSION**

In conclusion, a simple, compact and fully automated Seebeck coefficient measurement setup is developed for accurate measurement in the wide temperature range (80-650 K). The temperature gradient is precisely controllable during measurement and thus the measurement of Seebeck coefficient at desired average sample temperature can be easily obtained. A unique design of dipstick cryostat and sample holder allow the characterization of varieties of samples without any constraint on its geometrical shape, size and resistivity range, which is the most important feature of designed instrument with a low cost hardware. Desired temperature gradient ($\Delta T$) can be achieved with two heaters powered by dual channel power supply, and thus heat flow can be controlled from both side of the sample. System uses minimum instrumentation and measuring probes, which also minimize the hardware complexity and reduction in noise interferences within its own wires used. Calibration of thermocouples and absolute noise has been performed with the help of constantan, chromel and alumel wires, and measured data from our setup are found to be very accurate, and well consistent with earlier report. Seebeck coefficient of *n*-type GaN thin films has been determined and compared with available literature, which shows the versatile nature and capability of the setup for characterization of various type of thermoelectric materials.

**ACKNOWLEDGEMENTS**

A.K. would like to acknowledge the financial support received from the Department of Science and Technology, India through DST-INSPIRE Faculty scheme (DST/INSPIRE/04/2015/001572). The authors acknowledge Mr. Ramcharan Meena, S. K. Saini, D. K. Prabhakar and other workshop staff for their support in the fabrication process of the vacuum chamber and sample holder parts.